\documentclass{article}
\usepackage{spconf,amsmath,graphicx}
\usepackage{multirow}
\usepackage{booktabs}

\title{CTC Blank Triggered Dynamic Layer-skipping for Efficient CTC-based Speech Recognition}

\name{Junfeng Hou, Peiyao Wang, Jincheng Zhang, Meng Yang, Minwei Feng, Jingcheng Yin}
\address{Netease BizEase, Hangzhou, Zhejiang, China \\ \texttt{\{houjunfeng,fengminwei\}@corp.netease.com} }
\begin{document}
%
\maketitle
\begin{abstract}
Deploying end-to-end speech recognition models with limited computing resources remains challenging, despite their impressive performance. Given the gradual increase in model size and the wide range of model applications, selectively executing model components for different inputs to improve the inference efficiency is of great interest. In this paper, we propose a dynamic layer-skipping method that leverages the CTC blank output from intermediate layers to trigger the skipping of the last few encoder layers for frames with high blank probabilities. Furthermore, we factorize the CTC output distribution and perform knowledge distillation on intermediate layers to reduce computation and improve recognition accuracy. Experimental results show that by utilizing the CTC blank, the encoder layer depth can be adjusted dynamically, resulting in 29\% acceleration of the CTC model inference with minor performance degradation.
\end{abstract}
\section{Introduction}
\label{sec:intro}
End-to-end models (E2E) including connectionist temporal classification (CTC) \cite{graves2006connectionist}, recurrent neural network transducer (RNN-T) \cite{graves2012sequence} and attention-based encoder-decoder (AED) \cite{chan2016listen} have been widely used for automatic speech recognition (ASR). Benefiting from the E2E modeling property, the inference pipeline of these models are greatly simplified so that speech recognition can be readily performed
in many scenarios. However, these models usually require a large acoustic encoder \cite{whisper, pratap2023scaling} to achieve satisfying performance, which has a large impact on models' runtime efficiency.

\begin{figure}[htb]
		\centering		\centerline{\includegraphics[width=0.32\textwidth]{"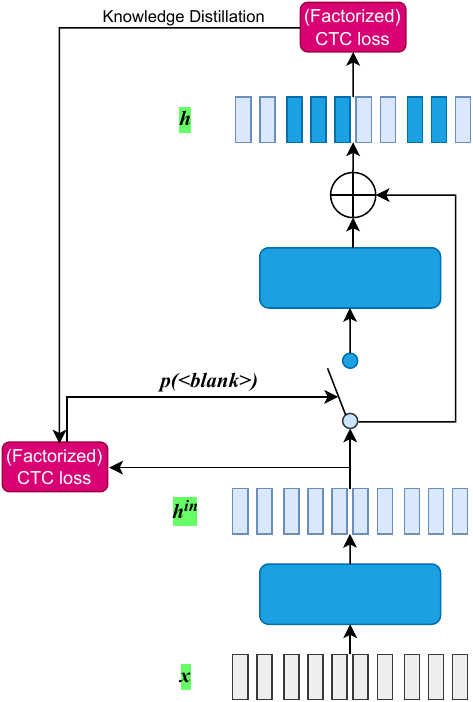"}}
	%
	\caption{Architecture of CTC blank triggered dynamic layer-skipping model. The encoder consists of two parts. The first part is consistently active, while the second part's execution is determined by CTC blank.}
	\label{fig:architecture}
\end{figure}

Model compression methods, such as distillation \cite{hinton2015distilling, chang2022distilhubert}, pruning \cite{lai2021parp}, and quantization \cite{tan2021compressing}, have been explored to reduce computation complexity and improve runtime efficiency. However, the compressed model has a fixed architecture for all inputs, which might be overqualified for some input. To address this issue, dynamic models \cite{han2021dynamic}, such as early exiting, layer-skipping \cite{lee21e_interspeech} and LayerDrop \cite{fan2019reducing}, have been proposed to selectively execute submodules for different inputs, achieving a better trade-off between performance and efficiency. For speech recognition, a dynamic encoder has been designed to process the beginning part of audio with a small encoder and the following part with a full-size encoder \cite{dynamicEncTransducer}. To achieve a more flexible dynamic model, fine-grained lightweight arbitrator networks \cite{Xie2022} or gate predictors \cite{i3d, alexandridis2023gated} have been proposed to decide whether to skip submodules.

For RNN-T and CTC models, a special blank symbol $\phi$ corresponding to ``output nothing'' is introduced. And previous works have found $\phi$ can guide the dynamic submodule skipping of RNN-T models \cite{ctc_guide, Tian2021FSRAT, yang2023blank, factorized_blk}. Inspired by these models, we propose a CTC-based dynamic layer-skipping method triggered by CTC blank for CTC-based models (refer to Fig. \ref{fig:architecture}), which conduct ``layer dropping/skipping'' in frame-level. We assume that frames assigned by blank are redundant and a shallow encoder is sufficient for acoustic modeling. To improve performance, we apply knowledge distillation to align the CTC output of the shallow encoder with that of the full-size encoder. To achieve faster inference speed, we propose factorized CTC to facilitate frame skipping in frame-synchronous CTC beam-search \cite{zhang2022wenet, chen2016phone}.

The main contributions of this paper are as follows:
\begin{itemize}
\item We verify the effectiveness of using CTC blank to trigger encoder layer-skipping in CTC-based models.
\item We demonstrate that applying knowledge distillation from the full-size encoder output to the intermediate layer can enhance the performance of dynamic layer-skipping encoder.
\item We propose factorized CTC to further improve runtime efficiency in combination with the frame skipping beam-search strategy.
\end{itemize}

\section{method}
\label{sec:layer-skipping}
\subsection{CTC}
\label{ssec:ctc}
The CTC \cite{graves2006connectionist} based E2E ASR model addresses the issue of length mismatch between acoustic frames and output token sequences by introducing an extra output blank token $\phi$ to allow output nothing but blank. For a sequence of input acoustic frames $\textbf{x}=(x_1, x_2, \cdots, x_T)$ of length T, the model's encoder produces hidden representation $\textbf{h}=(h_1, h_2, \cdots, h_{T'})$ where $T' < T$ because of downsampling. Then the model's decoder generate frame-wise output probabilities $\textbf{p}=(p_1, p_2, \cdots, p_{T'}$), corresponding to the vocabulary $V \cup \phi$. Given a sequence of output label $\textbf{y}=(y_1, y_2, \cdots, y_U)$ of length $U$, the loss function is defined as the probability of all possible alignments between $\textbf{x}$ and $\textbf{y}$:
\begin{align}
    p_t &= \text{softmax}(W h_t)\\
    L_{ctc}(y) &= \sum_{\pi\in \beta^{-1}(y)} {\log{p(\pi \mid x)}}
\end{align}
Where $\beta$ is a many-to-one mapping to remove repetitive labels and blank labels in an alignment $\pi$. During inference, a large proportion of acoustic frames are classified as blank frames, as shown in Fig. \ref{fig:ctc_blk}.

\subsection{CTC blank triggered dynamic layer-skipping}
\label{sssec:layer_skipping}

The architecture of our proposed CTC blank triggered dynamic layer-skipping model is illustrated in Fig. \ref{fig:architecture}. The input feature $\textbf{x}$ is transformed into an intermediate hidden representation $h^{in}$ through $K$ encoder layers. Subsequently, the CTC output probability derived from the intermediate layer, which includes blank and real tokens, is generated for layer-skipping. As blank represents no output for an input frame, the CTC blank probability $p_t^{in}(\phi)$ is used as a gating signal to trigger layer-skipping. Frames with a blank probability smaller than a predefined threshold $\tau$, referred to as significant frames, are processed continuously by the subsequent ($L$-$K$) encoder layers. Other frames are regarded as redundancy frames and are copied to the subsequent layers' output without calculation. The details of the calculation are shown in Equations (\ref{eqn:skip1}-\ref{eqn:skip3}).
\begin{align}
    \label{eqn:skip1}
    h^{in}_t&=Enc_1^K(x_t) \\
    \label{eqn:skip2}
    p^{in}_t&=\text{softmax}(W^{in} h^{in}_t) \\
    \label{eqn:skip3}
    h_t&= \begin{cases}
        h^{in}_t, & \text{if } p^{in}_{t:t-2}(\phi) > \tau \\
        Enc_{K+1}^L(h^{in}_t), & \text{else}
    \end{cases}
\end{align}

This procedure is based on the assumption that the alignment obtained from the intermediate layers is close to that of the final output layer. If this is not the case, the redundancy frames which are skipped ahead may play a crucial role in determining the final output as well, making the layer-skipping for such frames degrade the performance. To tackle this issue, we permit more frames to pass through the full-size encoder. Frame $t$ is skipped only when the blank probabilities of frame
t and previous 2 frames are larger than a threshold (See Equation (\ref{eqn:skip3}), denoted as ``spike extension''). Our experimental results in Section \ref{sec:spike_extend} confirm the efficacy of this trick.

\subsubsection{Multi-task CTC and knowledge distillation for improving performance}
\label{sssec:regularized_trick}

We also apply knowledge distillation (KL) to align the spike emitting between the intermediate layers and final output so as to avoid information loss caused by layer-skipping. Prior research has demonstrated that KL from distinct layers can facilitate layer pruning and frame skipping \cite{lee21e_interspeech, dynamicEncTransducer}. Meanwhile the inplace distillation technique \cite{inplaceDistill} is also proven to be effective for improving performance. Therefore, we adopt the same KL loss together with CTC losses:
\begin{align}
Loss &= L_{mtl} + \lambda \times \text{KL\_div}(p^{in}_t \: \Vert \: p_t) \\
L_{mtl} &= L_{ctc} + L_{ctc}^{in}
\end{align}
Here $L_{mtl}$ is the multi-task learning (MTL) loss, and $L_{ctc}^{in}$ and $L_{ctc}$ are the CTC losses of the intermediate encoder and the full-size encoder. $\text{KL\_div}$ is the KL-divergence loss, and $p^{in}_t$ and $p_t$ denote the output probabilities of the intermediate encoder and the full-size encoder, respectively. $\lambda$ is set to 0.5.
\subsubsection{Factorized CTC for improving efficiency}
\label{sssec:factorized_ctc}

\begin{figure*}[htp]
        \includegraphics[width=0.93\columnwidth]{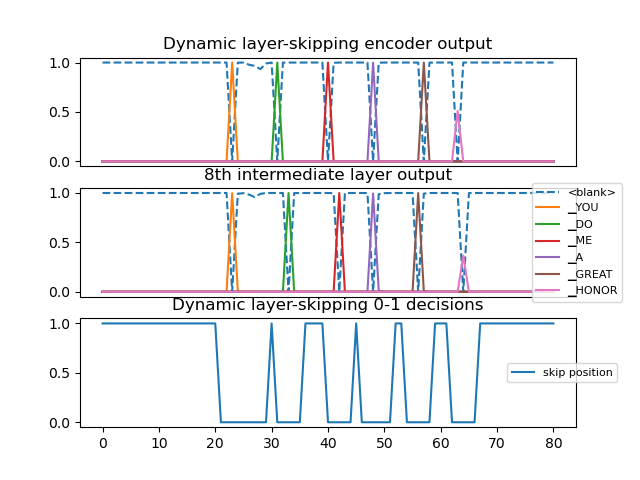}
        \label{fig:figure1}
        \includegraphics[width=0.93\columnwidth]{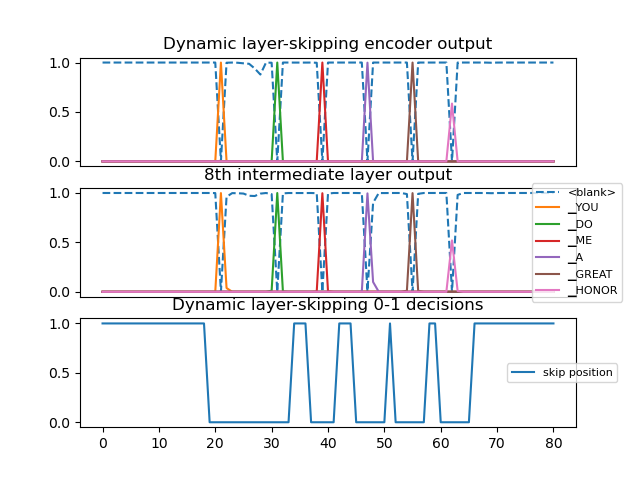}
        \label{fig:figure2}
        \vspace{0.0cm}
    \caption{The CTC output of different layer-skipping models. (Left): Model without KL loss; (Right): Model with KL loss.}
    \label{fig:ctc_blk}
\end{figure*}

The Frame skipping decoding strategy is commonly utilized to accelerate CTC beam search by disregarding frames with higher blank probabilities, which have little impact on the beam search process \cite{lee21e_interspeech, dynamicEncTransducer}. Building on this idea, we introduce a factorized CTC output distribution for intermediate and final layers to accelerate the layer-skipping procedure.

Following the approach proposed in \cite{factorized_blk}, we ignore the probabilities of non-blank tokens in significant frames of intermediate layers, and recalculate the probabilities based on the full-size encoder. Additionally, frames with high blank probabilities from the full-size encoder can also be skipped, further hastening the CTC beam search process. The factorized distributions are defined as follows:
\begin{align}
    p^b_t&=\text{sigmoid}(v^{T}h_t) \\
    p^{nb}_t&=(1-p_t^b) \cdot \text{softmax}(W^{nb}h_t) \\
    p_t&=\text{concat}(p^b;p^{nb})
\end{align}
Here $v$ is a vector and $W^{nb}$ maps $h_t$ to $V$ dimension.

\begin{table}[hbp]
  \centering
  \caption{WER and RTF comparisons between the baseline and our proposed layer-skipping models on librispeech \textbf{test-clean/test-other} splits. ``MTL'' means multi-task learning CTC losses $L_{mtl}$. ``KL'' means knowledge distillation loss to intermediate layers. ``CTC'' and ``Rescore'' refer to decoding modes of CTC beam-search and attention-based rescoring.}
  \label{tab:model_performance}
  \begin{tabular}{c|c|c|c}
    \toprule
    \textbf{Model} & \textbf{Mode} & {\textbf{WER}} & \textbf{RTF} \\
    \midrule
    Baseline & \multirow{4}{*}{CTC} & 3.76/9.50 & 0.078 \\
    \cline{1-1} \cline{3-4}
    Baseline + MTL & & 3.65/9.35 & 0.078 \\
    \cline{1-1} \cline{3-4}
    Layer-skipping(our) & & 3.88/9.54 & 0.048  \\
    \cline{1-1} \cline{3-4}
    Layer-skipping(our) + KL &  & 3.70/9.34 & 0.048 \\
    \hline
    Baseline & \multirow{4}{*}{Rescore} & 3.32/8.67  & 0.086 \\
    \cline{1-1} \cline{3-4}
    Baseline + MTL & & 3.23/8.51 & 0.086 \\
    \cline{1-1} \cline{3-4}
    Layer-skipping(our) & & 3.42/8.74 & 0.065  \\
    \cline{1-1} \cline{3-4}
    Layer-skipping(our) + KL & & 3.23/8.55 & 0.065 \\
    \bottomrule
  \end{tabular}
\end{table}

\section{Experiments and Results}
\label{sec:result}

We evaluate the proposed method on the LibriSpeech dataset \cite{panayotov2015librispeech}, which consists of 960 hours of labeled audio training data. The models' performance in terms of Word Error Rate (WER) and Real-Time Factor (RTF) is evaluated on the test-clean and test-other splits. We adopt the ``Conformer U2++'' recipe\footnote{https://github.com/wenet-e2e/wenet/tree/main/examples/librispeech/s0} from WeNet \cite{zhang2022wenet}, where N-best hypotheses are generated by CTC beam-search and then rescored by attention based E2E model.

\begin{table*}[htp]
  \centering
  \caption{Performance of dynamic layer-skipping at different encoder positions.}
  \label{tab:different_KL_layer}
  \begin{tabular}{c|c|c|c|c|c}
    \toprule
    \textbf{ Model} & \textbf{ skip layer idx} & \textbf{spike extension} & \textbf{WER} & \textbf{RTF} & \textbf{skip ratio} \\
    \midrule
    baseline-11 layer & - & - & 3.47/8.86 & 0.070 & 0\% \\
    \hline
    \multirow{6}{*}{our} & \multirow{2}{*}{4} & Y & 3.36/8.87 & 0.047 & 27.98\% \\
   \cline{3-6}
    &  & N & 3.76/9.23 & 0.037 & 47.67\% \\
   \cline{2-6}
     & \multirow{2}{*}{6} & Y & 3.31/8.70 & 0.054 & 33.38\% \\
    \cline{3-6}
     & & N & 3.61/9.05 & 0.043 & 49.26\% \\
    \cline{2-6}
     & \multirow{2}{*}{8} & Y & 3.23/8.55 & 0.065 & 33.89\% \\
    \cline{3-6}
     &  & N & 3.53/8.75 & 0.055 & 52.89\% \\ 
    \bottomrule 
  \end{tabular}
\end{table*}

In all of our experiments, we extract 80-dimensional log Mel-filter bank features, which are then processed by a convolution sub-sampling module with a total stride of 4. The features are fed to a 12-layer Conformer encoder with a hidden size of 256. Both the CTC and attention rescoring decoders share the encoder's hidden representation and output 5000 Byte Pair Encoding (BPE) tokens \cite{sennrich2015neural}. The attention rescoring decoders consist of a forward and a backward 3-layer Transformer decoder, each with a hidden size of 256. All other configurations are the same as those in the recipe. We set the blank threshold $\tau$ to 0.99 for layer-skipping and frame skipping during inference.

\subsection{Performance of dynamic layer-skipping model}
\label{sec:spike_extend}

The main results are presented in Table \ref{tab:model_performance}. We conduct layer-skipping and MTL CTC loss at layer 8, which is found to be the optimal choice (See Table \ref{tab:different_KL_layer}). We compare the effectiveness of the layer-skipping model with the baseline U2++ model for both CTC beam-search and attention rescoring decoding methods. Compared to the baseline, the proposed layer-skipping model achieves a 38\% and 23\% reduction in RTF for the two decoding modes, respectively, while maintaining similar WERs to the baseline model. As MTL loss improves the baseline performance \cite{lee21e_interspeech}, we also apply the MTL loss for the baseline model and find that the proposed model performs worse than baseline with MTL loss. We hypothesize that the alignment mismatch between the intermediate and final layers results in the performance loss. Therefore, we conduct a KL loss for the layer-skipping model and achieve similar performance to the MTL baseline while maintaining the inference speedup. We also evaluate the 8th intermediate layer by-product output from the ``Layer-skipping + KL'' model to show the need of more layers for significant frames. The ``Rescore'' mode's test-clean/test-other WERs is 3.75/9.76, which is much worse than the whole dynamic layer-skipping model. Since the rescoring mode yields better performance, we only report the rescoring results in the following experiments.

The alignment figure is presented in Fig. \ref{fig:ctc_blk}. As shown, most blank frames are skipped, and only frames around spiking ones are processed by the full-size encoder. The intermediate layer tends to emit tokens later than the final encoder, allowing it to collect more information from contextual frames. However, this also means that the $30{th}$ frame is not fed to the final layer, which may harm the performance. After introducing the KL loss, the spikes of the intermediate and final layers are well aligned, benefiting the layer-skipping.

We also evaluate the dynamic layer-skipping position in Table \ref{tab:different_KL_layer} and find a tradeoff between RTFs and WERs. As layer skipping is conducted in later layers, the skip ratio increases. Additionally, spike extension is important for performance. After the extension, the skip ratio is 33\%, which means the average ``layer number'' is about 10.68 (0.33*8+0.67*12=10.68). Therefore, we conduct a small baseline model of 11 layers and find that the proposed model has better performance while maintaining comparable RTFs.

\begin{table}[h]
  \centering
  \caption{Performance of dynamic layer-skipping with factorized CTC.}
  \label{tab:factorized_ctc}
  \begin{tabular}{c|c|c|c}
    \toprule
    \textbf{ Model} & \textbf{ CTC type} & \textbf{WER} & \textbf{RTF} \\
    \midrule
    Baseline + MTL & CTC & 3.23/8.51 & 0.086 \\
    \hline
    \multirow{2}{*}{Our} & CTC & 3.23/8.55 & 0.065 \\
   \cline{2-4}
    & factorized CTC & 3.29/8.61 & 0.061 \\
    \bottomrule
  \end{tabular}
\end{table}

Furthermore, we replace the CTC loss with factorized CTC for both intermediate layer and output layer. The results are shown in Table \ref{tab:factorized_ctc}. Together with the frame-skipping during inference, another 6\% RTFs reduction is achieved with minor performance degradation.

\section{conclusion}
\label{sec:conclusion}
In this paper, we introduce a CTC blank-triggered dynamic layer-skipping method to skip the computation of the last few encoder layers for frames with high blank probabilities. We factorize the CTC output distribution and perform knowledge distillation on intermediate layers to reduce computation and improve recognition accuracy. We also present the performance of different skipping positions and visualize the blank skipping cases. In the future, we plan to explore more efficient skipping methods based on other alignment mechanisms.



\begin{thebibliography}{10}

\bibitem{graves2006connectionist}
Alex Graves, Santiago Fern{\'a}ndez, Faustino Gomez, and J{\"u}rgen
  Schmidhuber,
\newblock ``Connectionist temporal classification: labelling unsegmented
  sequence data with recurrent neural networks,''
\newblock in {\em Proceedings of the 23rd international conference on Machine
  learning}, 2006, pp. 369--376.

\bibitem{graves2012sequence}
Alex Graves,
\newblock ``Sequence transduction with recurrent neural networks,''
\newblock {\em arXiv preprint arXiv:1211.3711}, 2012.

\bibitem{chan2016listen}
William Chan, Navdeep Jaitly, Quoc Le, and Oriol Vinyals,
\newblock ``Listen, attend and spell: A neural network for large vocabulary
  conversational speech recognition,''
\newblock in {\em 2016 IEEE international conference on acoustics, speech and
  signal processing (ICASSP)}. IEEE, 2016, pp. 4960--4964.

\bibitem{whisper}
Alec Radford, Jong~Wook Kim, Tao Xu, Greg Brockman, Christine McLeavey, and
  Ilya Sutskever,
\newblock ``Robust speech recognition via large-scale weak supervision,''
\newblock in {\em International Conference on Machine Learning}. PMLR, 2023,
  pp. 28492--28518.

\bibitem{pratap2023scaling}
Vineel Pratap, Andros Tjandra, Bowen Shi, Paden Tomasello, Arun Babu, Sayani
  Kundu, Ali Elkahky, Zhaoheng Ni, Apoorv Vyas, Maryam Fazel-Zarandi, et~al.,
\newblock ``Scaling speech technology to 1,000+ languages,''
\newblock {\em arXiv preprint arXiv:2305.13516}, 2023.

\bibitem{hinton2015distilling}
Geoffrey Hinton, Oriol Vinyals, and Jeff Dean,
\newblock ``Distilling the knowledge in a neural network,''
\newblock {\em arXiv preprint arXiv:1503.02531}, 2015.

\bibitem{chang2022distilhubert}
Heng-Jui Chang, Shu-wen Yang, and Hung-yi Lee,
\newblock ``Distilhubert: Speech representation learning by layer-wise
  distillation of hidden-unit bert,''
\newblock in {\em ICASSP 2022-2022 IEEE International Conference on Acoustics,
  Speech and Signal Processing (ICASSP)}. IEEE, 2022, pp. 7087--7091.

\bibitem{lai2021parp}
Cheng-I~Jeff Lai, Yang Zhang, Alexander~H Liu, Shiyu Chang, Yi-Lun Liao,
  Yung-Sung Chuang, Kaizhi Qian, Sameer Khurana, David Cox, and Jim Glass,
\newblock ``Parp: Prune, adjust and re-prune for self-supervised speech
  recognition,''
\newblock {\em Advances in Neural Information Processing Systems}, vol. 34, pp.
  21256--21272, 2021.

\bibitem{tan2021compressing}
Ke~Tan and DeLiang Wang,
\newblock ``Compressing deep neural networks for efficient speech
  enhancement,''
\newblock in {\em ICASSP 2021-2021 IEEE International Conference on Acoustics,
  Speech and Signal Processing (ICASSP)}. IEEE, 2021, pp. 8358--8362.

\bibitem{han2021dynamic}
Yizeng Han, Gao Huang, Shiji Song, Le~Yang, Honghui Wang, and Yulin Wang,
\newblock ``Dynamic neural networks: A survey,''
\newblock {\em IEEE Transactions on Pattern Analysis and Machine Intelligence},
  vol. 44, no. 11, pp. 7436--7456, 2021.

\bibitem{lee21e_interspeech}
Jaesong Lee, Jingu Kang, and Shinji Watanabe,
\newblock ``{Layer Pruning on Demand with Intermediate CTC},''
\newblock in {\em Proc. Interspeech 2021}, 2021, pp. 3745--3749.

\bibitem{fan2019reducing}
Angela Fan, Edouard Grave, and Armand Joulin,
\newblock ``Reducing transformer depth on demand with structured dropout,''
\newblock in {\em International Conference on Learning Representations}, 2020.

\bibitem{dynamicEncTransducer}
Yangyang Shi, Varun Nagaraja, Chunyang Wu, Jay Mahadeokar, Duc Le, Rohit
  Prabhavalkar, Alex Xiao, Ching-Feng Yeh, Julian Chan, Christian Fuegen, Ozlem
  Kalinli, and Michael Seltzer,
\newblock ``Dynamic encoder transducer: A flexible solution for trading off
  accuracy for latency,''
\newblock 08 2021, pp. 2042--2046.

\bibitem{Xie2022}
Yi~Xie, Jonathan Macoskey, Martin Radfar, Feng-Ju~(Claire) Chang, Brian King,
  Ariya Rastrow, Athanasios Mouchtaris, and Grant Strimel,
\newblock ``Compute cost amortized transformer for streaming asr,''
\newblock in {\em Interspeech 2022}, 2022.

\bibitem{i3d}
Yifan Peng, Jaesong Lee, and Shinji Watanabe,
\newblock ``I3d: Transformer architectures with input-dependent dynamic depth
  for speech recognition,''
\newblock in {\em ICASSP 2023 - 2023 IEEE International Conference on
  Acoustics, Speech and Signal Processing (ICASSP)}, 2023, pp. 1--5.

\bibitem{alexandridis2023gated}
Anastasios Alexandridis, Kanthashree~Mysore Sathyendra, Grant~P Strimel,
  Feng-Ju Chang, Ariya Rastrow, Nathan Susanj, and Athanasios Mouchtaris,
\newblock ``Gated contextual adapters for selective contextual biasing in
  neural transducers,''
\newblock in {\em ICASSP 2023-2023 IEEE International Conference on Acoustics,
  Speech and Signal Processing (ICASSP)}. IEEE, 2023, pp. 1--5.

\bibitem{ctc_guide}
Yongqiang Wang, Zhehuai Chen, Chengjian Zheng, Yu~Zhang, Wei Han, and Parisa
  Haghani,
\newblock ``Accelerating rnn-t training and inference using ctc guidance,''
\newblock in {\em ICASSP 2023 - 2023 IEEE International Conference on
  Acoustics, Speech and Signal Processing (ICASSP)}, 2023, pp. 1--5.

\bibitem{Tian2021FSRAT}
Zhengkun Tian, Jiangyan Yi, Ye~Bai, Jianhua Tao, Shuai Zhang, and Zhengqi Wen,
\newblock ``Fsr: Accelerating the inference process of transducer-based models
  by applying fast-skip regularization,''
\newblock in {\em Interspeech}, 2021.

\bibitem{yang2023blank}
Yifan Yang, Xiaoyu Yang, Liyong Guo, Zengwei Yao, Wei Kang, Fangjun Kuang, Long
  Lin, Xie Chen, and Daniel Povey,
\newblock ``Blank-regularized ctc for frame skipping in neural transducer,''
\newblock {\em arXiv preprint arXiv:2305.11558}, 2023.

\bibitem{factorized_blk}
Duc Le, Frank Seide, Yuhao Wang, Yang Li, Kjell Schubert, Ozlem Kalinli, and
  Michael~L. Seltzer,
\newblock ``Factorized blank thresholding for improved runtime efficiency of
  neural transducers,''
\newblock in {\em ICASSP 2023 - 2023 IEEE International Conference on
  Acoustics, Speech and Signal Processing (ICASSP)}, 2023, pp. 1--5.

\bibitem{zhang2022wenet}
Binbin Zhang, Di~Wu, Zhendong Peng, Xingchen Song, Zhuoyuan Yao, Hang Lv, Lei
  Xie, Chao Yang, Fuping Pan, and Jianwei Niu,
\newblock ``Wenet 2.0: More productive end-to-end speech recognition toolkit,''
\newblock {\em arXiv preprint arXiv:2203.15455}, 2022.

\bibitem{chen2016phone}
Zhehuai Chen, Wei Deng, Tao Xu, and Kai Yu,
\newblock ``Phone synchronous decoding with ctc lattice.,''
\newblock in {\em Interspeech}, 2016, pp. 1923--1927.

\bibitem{inplaceDistill}
Jiahui Yu and Thomas Huang,
\newblock ``Universally slimmable networks and improved training techniques,''
\newblock in {\em 2019 IEEE/CVF International Conference on Computer Vision
  (ICCV)}, 2019, pp. 1803--1811.

\bibitem{panayotov2015librispeech}
Vassil Panayotov, Guoguo Chen, Daniel Povey, and Sanjeev Khudanpur,
\newblock ``Librispeech: an asr corpus based on public domain audio books,''
\newblock in {\em 2015 IEEE international conference on acoustics, speech and
  signal processing (ICASSP)}. IEEE, 2015, pp. 5206--5210.

\bibitem{sennrich2015neural}
Rico Sennrich, Barry Haddow, and Alexandra Birch,
\newblock ``Neural machine translation of rare words with subword units,''
\newblock {\em arXiv preprint arXiv:1508.07909}, 2015.

\end{thebibliography}

\end{document}